\documentclass[12pt]{article}
\usepackage[latin1]{inputenc}
\usepackage{amsfonts}
\usepackage{amssymb}
\usepackage{amsmath}
\usepackage{latexsym}
\usepackage{graphicx}

\begin{document}
	\title{Predicting the evolution of the COVID-19 epidemic with the A-SIR model: Lombardy, Italy and S\~ao Paulo state, Brazil}
	\author{Armando G. M. Neves$^{1,3}$ and Gustavo Guerrero$^{2,3}$
		\\
		\normalsize{$^{1}$Departamento de Matem\'atica, Universidade Federal de Minas Gerais}
		\\ 	\normalsize{aneves@mat.ufmg.br}\\
		\normalsize{$^{2}$Departamento de F\'isica, Universidade Federal de Minas Gerais}
		\\ 	\normalsize{guerrero@fisica.ufmg.br}\\
		\normalsize{$^{3}$COVID-19 Modeling Task Force, }\\ \normalsize{Minas Gerais, Brazil}
	}
	
	\date{\today}
	\maketitle
	\begin{abstract}
		The presence of a large number of infected individuals with few or no symptoms is an important epidemiological difficulty and the main mathematical feature of COVID-19. The A-SIR model, i.e. a SIR (Susceptible-Infected-Removed) model with a compartment for infected individuals with no symptoms or few symptoms was proposed by Giuseppe Gaeta, arXiv:2003.08720 [q-bio.PE] (2020). In this paper we investigate a slightly generalized version of the same model and propose a scheme for fitting the parameters of the model to real data using the time series only of the deceased individuals. The scheme is applied to the concrete cases of Lombardy, Italy and S\~ao Paulo state, Brazil, showing  different aspects of the epidemics. For each case we show that we may have good fits to the 
		data up to the present, but with very large differences in the future behavior. The reasons behind such disparate outcomes are the uncertainty on the value of a key parameter, the probability that an infected individual is fully symptomatic, and on the intensity of the social distancing measures adopted. This conclusion enforces the necessity of trying to determine the real number of infected individuals in a population, symptomatic or asymptomatic.    
	\end{abstract}

	
	\section{Introduction}
	Although there are good models for predicting the time evolution of an epidemic of diseases such as influenza or measles, models of the same type are not working for the COVID-19. An important feature of the COVID-19 is that it may be asymptomatic or mildly symptomatic in some patients, although causing severe respiratory symptoms in others. patients. As a consequence, there is a large number of undocumented infections \cite{Li}.
	
	The lack of tests for assessing the health state of large samples of the populations contributes to the spread of the COVID-19, 
	as asymptomatic individuals may not isolate themselves. Although there is no clear distinction between symptomatic and asymptomatic, in this paper we will use the acronym MSA (mildly symptomatic or asymptomatic) to mean a case of COVID-19 weak enough for not causing 
	death or lead to hospitalization and
	probably unreported due to the lack of tests.
	
	We consider that a good step towards a good predicting model for COVID-19 has been taken in \cite{Gaeta}. In Sect. \ref{secasir} we will describe a slight generalization of the A-SIR model proposed in that work and use it in the rest of this paper in predicting the possible evolution of the COVID-19 epidemics in Lombardy, Italy and S\~ao Paulo state, Brazil. 
	
	The A-SIR model is just the traditional SIR (Susceptible-Infected-Removed) model for epidemics, introduced by Kermack and McKendrick \cite{kmk} almost 100 years ago, with one extra compartment for accounting the MSA infected individuals. The MSA can still transmit the disease to susceptible individuals, but, as they mostly ignore their condition, it is reasonable that they will remain for larger periods transmitting the disease when compared to fully symptomatic individuals, which will probably isolate themselves after a few days. Of course, when the epidemic is already well developed, a large fraction of the population may have been MSA infected, and, as these individuals are healed, they will contribute to largely decrease the number of susceptible people.
	
	Although the proportion of symptomatic cases has already been estimated to be $16\%$ for the development of the disease in China \cite{Li}, 
	we will take the liberty to explore the possibility that this proportion may be larger or smaller.
	
	As a support to the possibility that there are less symptomatic cases than previously estimated, we cite one among the many reports by the Imperial College COVID-19 response team \cite{imperial13}. Referring to 11 European countries, the report states that ``In all countries, we estimate there are orders of magnitude fewer infections detected than true infections, mostly likely due to mild and asymptomatic infections as well as limited testing capacity". Figure 2 in that report illustrates that.
	
	Supporting the other possibility, Lavezzo et al. \cite{lavezzo} state that at V\`o, Italy, the asymptomatic cases were a fraction of $43.2\%$ of the total. Although clearly casting some doubt, we also cite \cite{iceland}. The paper states ``Among the participants with positive results for SARS-CoV-2, symptoms of Covid-19 were reported (...) by $57\%$ of those in the overall population-screening group. However, $29\%$ of participants who tested negative in the overall population-screening group also reported having symptoms".

	One reason for the uncertainty in the outcome of mathematical models for COVID-19 is that the models usually contain parameters for which reasonable values are taken, but sometimes without full scientific support. In particular, the models are extremely sensitive to the infection rate $\beta_0$, see (\ref{asirbasic}). We will show in this paper how to ignore the data on the number of currently infected people. These are prone to a large uncertainty, because of the MSA cases, but also underreporting of the symptomatic cases due to the lack of tests. We will use only the data on the number of deaths due to the COVID-19, expected to be more faithful. We will restrict for the time being to the study of the development of COVID-19 in
	Lombardy and in the state of S\~ao Paulo. Both cases result in good fits of the model to the data. It will turn out that an important part of the fitting procedure is the way of tuning  
	the value of the infection rate $\beta_0$ to the data.
	
	An important question is what will happen when the social distancing measures currently in act in most countries are relaxed. One bad possibility is that a second wave of COVID-19 will arise. If not mitigated, the potential number of deaths in the second wave may be larger than the deaths in the first wave. Another possibility is that sufficient herd immunity will have been acquired by the populations after the present epidemic and no large increase of cases should happen after relaxation of the social distancing. 
	
	We will show in this paper that neither of the above possibilities can be ruled out for Lombardy. Part of our ignorance is due to the fact that one key parameter of the A-SIR model, the probability that a newly infected individual is symptomatic, is still largely unknown. Another reason for not being able to predict the future of the epidemic is that we do not know how much the social distancing measures adopted were effective in reducing the infection rate of the model.  
	
	In the case of S\~ao Paulo state, Brazil, the fraction of deaths up to now is much smaller than in Lombardy. Although this is good, it also means that the population is still very susceptible. Strong economic pressure is being exerted on politicians for relaxation of the social distancing measures. We predict that even in the best of the possibilities, the number of infected individuals will steadily grow for a large period
	and in its peak it will be much larger than present values. Thus, social distancing measures should not be relaxed before the number of infected individuals is small. 
	We see that the increase in the number of cases may be catastrophic if social distancing measures are not strengthened.
	
	The paper is organized as follows. Sect. \ref{secasir} starts with a mathematical description of the model and all its parameters. Then we will talk about the linear regime, i.e. the behavior of the solutions of the model for a short time after the beginning of the epidemic. Finally, we will describe conditions for the population fraction of infected individuals to decrease and also the limit behavior after the epidemic is finished. Sect. \ref{secfit} describes the procedure for finding values of the parameters such that the output of the A-SIR model fits well the deaths data, both in Lombardy and in S\~ao Paulo. The paper is closed by Sect. \ref{secconc}, in which we draw some conclusions on the results obtained,
	and by Sect. \ref{secps}, in which we account for some changes in the conclusions because of 
	new data released during the time the paper was being written.

	\section{The A-SIR model}\label{secasir}
	Let $S(t)$, $I(t)$, $A(t)$ and $R(t)$ be the population fractions at time $t$ respectively of susceptible,  symptomatic infected, MSA infected and removed individuals. By susceptible, we mean individuals which were not yet infected by the SARS-Cov-2 virus. By symptomatic we mean \textit{fully} symptomatic individuals and by MSA we mean individuals which have either \textit{no} symptoms, or \textit{few} symptoms. By removed we mean individuals which were either healed after infection, or deceased. The fraction of removed individuals is composed by the sum of symptomatic removed $R_s(t)$
	and MSA removed individuals $R_a(t)$, according to whether the individuals were fully symptomatic before removal, or had mild or no symptoms. 
	The time span we are going to consider is of a few months, thus we may ignore births and deaths by reasons other than infection. In particular, we suppose that MSA or susceptible individuals do not die and that all infected individuals do not become susceptible again, at least for the time span we are considering.
	
	The A-SIR model is described by the following set of nonlinear ordinary differential equations:
	\begin{equation}\label{asireqs}
	\left\{ 
	\begin{array}{rcl}
	S'(t)&=& -\beta_0 S(I+\mu A)\\
	I'(t)&=& \beta_0 \xi S(I+\mu A) - \gamma_s I\\
	A'(t)&=& \beta_0 (1-\xi) S(I+\mu A) - \gamma_a A\\
	R_s'(t)&=& \gamma_s I\\
	R_a'(t)&=& \gamma_a A 
	\end{array}\right. \;.
	\end{equation}
	The latter two equations are not essential for solving the system. In fact, we may calculate $R_s$ and $R_a$ simply by integrating respectively $I(t)$ and $A(t)$. Another simple property of the model, proved by summing all the 5 equations, is that the sum of the fractions $S$, $I$, $A$, $R_s$ and $R_a$ is a constant. If we take $\mu=1$, this is exactly the same model as in \cite{Gaeta}, although with different notation. 
	
	All parameters above are considered to be positive and are interpreted as follows:
	\begin{itemize}
		\item $\beta_0$ is the infection rate of symptomatic individuals;
		\item $\mu \in (0,1]$ is a reduction factor such that the infection rate for the MSA is $\mu \beta_0$;
		\item $\xi \in (0,1)$ is the probability that a new infection event leads to a symptomatic case;
		\item $\gamma_s$ and $\gamma_a$ are respectively the inverses of the mean time symptomatic and MSA individuals remain infective.
		We suppose that $\gamma_s> \gamma_a$.
	\end{itemize}
	
	The mean removal time for symptomatic individuals will be considered to be around a week. This does not mean that individuals with symptoms will be healed after one week, but that these individuals, after showing symptoms for some days will either be hospitalized, or stay isolated at home. We will take then $\gamma_s=1/7 \, (\mathrm{days})^{-1}$. Following \cite{Gaeta}, we will take $\gamma_a=1/21 \, (\mathrm{days})^{-1}$, meaning that MSA individuals will remain active in the population for a larger time than symptomatic individuals. 
	
	We believe that we may take values such as above for $\gamma_s$ and $\gamma_a$ without risk of overestimating or underestimating the size of the epidemics. Other values could have been considered, but would not change the main conclusions. In the following we will explain how to use mortality data to infer the value of the contact rate $\beta_0$. It will turn out that the value of the parameter $\mu$ will not alter almost anything in the numeric predictions. On the contrary, we will see that the remaining parameter $\xi$ alters drastically the outcome of the model in the future.
	
\begin{figure}
		\begin{center}
			\includegraphics[width= 0.7\textwidth]{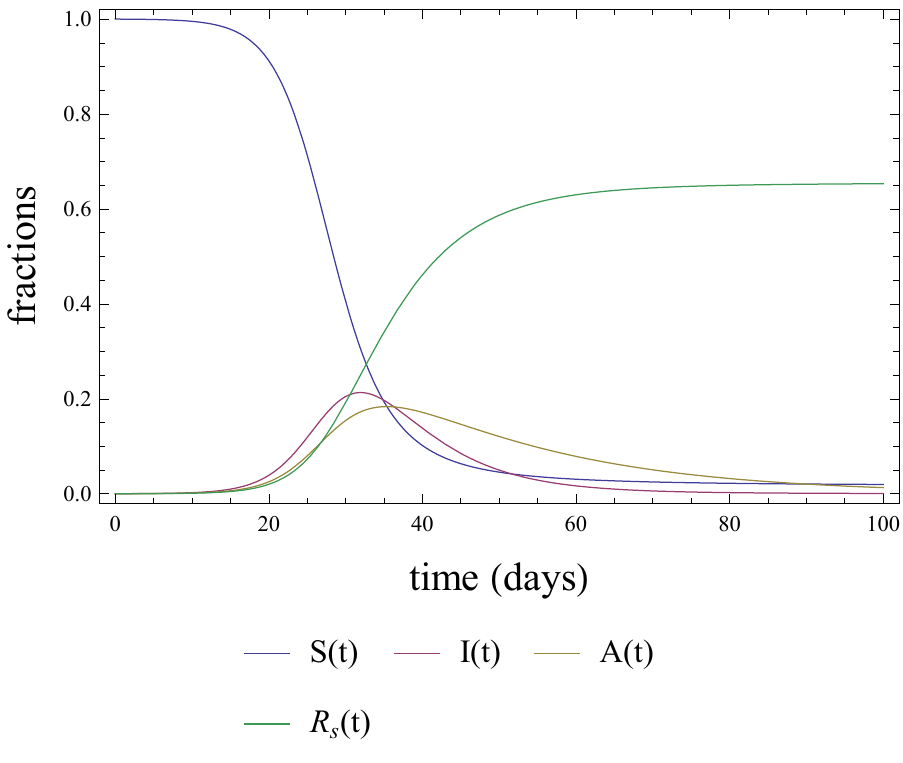}
			\caption{\label{asirbasic} Typical behavior in the A-SIR model of the fractions of susceptible, symptomatic infected, MSA infected and symptomatic removed individuals. Parameter values: $\beta_0=0.5$, $\mu=0.5$, $\xi=2/3$, $\gamma_s=1/7$, $\gamma_a=1/21$. The initial conditions are $S(0)=1$, $I(0)=A(0)= 0.0001$, $R(0)=0$.}
		\end{center}
	\end{figure}

	As a preparation for understanding things to come, we show in Fig. \ref{asirbasic} a typical graph of the fractions $S$, $I$, $A$ and $R_s$ as functions of time for a seemingly reasonable choice of parameters. The graphs are obtained by numerically solving Eqs.~(\ref{asireqs}).
	
	For this choice of parameters, note that about $65\%$ of the population are symptomatic removed 100 days after the start of the epidemic.
	Another thing to notice in the graphs of Fig. \ref{asirbasic} is that, although we used $\xi=2/3$ 
	expecting to obtain that
	2/3 of the cases are symptomatic, this 
	does not happen. In fact, by time $t=40$ the fraction of MSA is larger than the number of symptomatic individuals, even considering that
	the probability of a case being symptomatic 
	is larger than the probability of a MSA case.  The reason for obtaining a large number of MSA individuals 
	is not related to $\xi$, but to the fact that the mean time $\gamma_a^{-1}$ that an individual takes as MSA is larger than the mean time $\gamma_s^{-1}$ taken by
	a symptomatic individual.
	\begin{figure}
		\begin{center}
			\includegraphics[width= 0.7\textwidth]{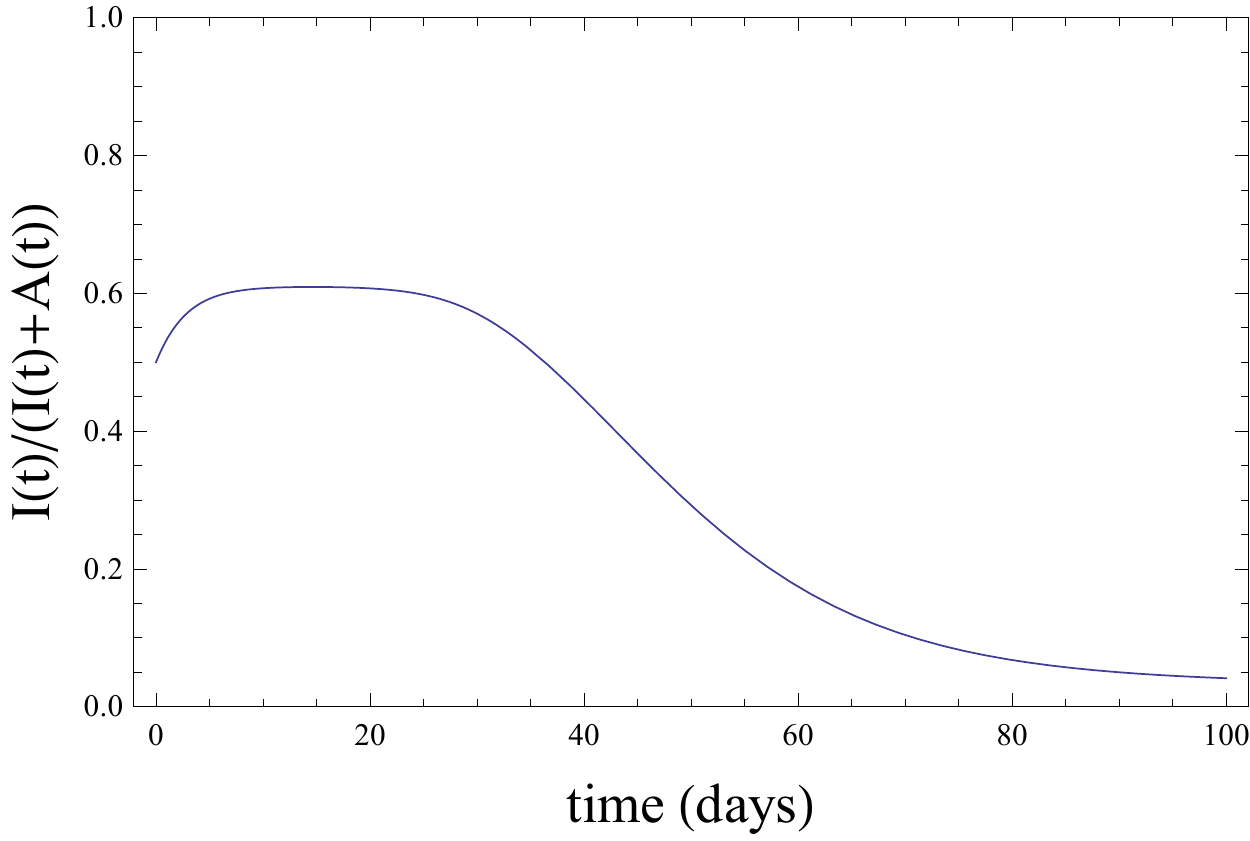}
			\caption{\label{ratiograph} The fraction of symptomatic cases as a function of time in the A-SIR model for the same parameter values and initial conditions as in Fig. \ref{asirbasic}.}
		\end{center}
	\end{figure}

	To better understand an important issue, in Fig. \ref{ratiograph} we plot, for the same set of parameters and initial conditions, the ratio $I(t)/(I(t)+A(t))$ of symptomatic to total cases. The figure
	shows that,  
	because of the initial condition $I(0)=A(0)$,
	the ratio of symptomatic to total infected individuals starts equal to 1/2. After a transient, it becomes almost constant around 0.6, not 2/3, and then decays to 0. 
	Part of this behavior will be explained in the next subsection.
	
	\subsection{The linear regime}\label{sublinear}
	Another feature of the A-SIR model, already noticed in \cite{Gaeta}, is that the fraction $S(t)$ is very well approximated by 1 for the initial times. We may use this to approximate the solution of Eqs.~(\ref{asireqs}) for small times. Substituting $S(t)$ by 1 in (\ref{asireqs}), we get
	\begin{equation}\label{linasireqs}
	\left\{ 
	\begin{array}{rcl}
	I'(t)&=& (\beta_0 \xi - \gamma_s) I+\beta_0 \xi \mu A \\
	A'(t)&=& \beta_0 (1-\xi) I+ (\mu \beta_0 (1-\xi) -\gamma_a) A
	\end{array}\right. \;,
	\end{equation}
	which is a linear system of ordinary differential equations with constant coefficients. Although the exact system  (\ref{asireqs}) cannot be exactly solved, its linear approximation for initial times can be solved in terms of the eigenvalues and eigenvectors of its coefficient matrix
	\begin{equation}\label{coeffmatrix}
	M \,=\, \left( \begin{array}{cc}
	\beta_0 \xi - \gamma_s & \beta_0 \xi \mu \\
	\beta_0 (1-\xi)& \mu \beta_0 (1-\xi) -\gamma_a
	\end{array} \right)\;.
	\end{equation}
	As $M$ is $2 \times 2$, its eigenvalues can be easily calculated as roots of a quadratic polynomial. It can be shown that the eigenvalues of $M$ are always real and that the smaller of them, denoted $\lambda_-$, is negative. The larger eigenvalue of $M$ will be denoted $\lambda_+$ and is positive, provided that $\beta_0$ is not too small, as will be seen ahead.
	Although the formula for $\lambda_+$ in terms of the parameters of the model is somewhat large, we may invert it and find a rather simple formula for $\beta_0$ as a function of $\lambda_+$ and the remaining parameters:
	\begin{equation} \label{betalambda}
	\beta_0 \,=\, \frac{(\lambda_+ + \gamma_s)(\lambda_+ + \gamma_a)}{(\lambda_+ + \gamma_a)\xi+ (\lambda_+ + \gamma_s) \mu (1-\xi) }\;.
	\end{equation}
	In Sect. \ref{secfit}, we will use the above formula along with an estimate of $\lambda_+$ derived from the data to fix the parameter $\beta_0$.
	
	A straightforward, but lengthy, calculation shows that for any initial conditions the solution of Eqs.~(\ref{linasireqs}) satisfies
	\begin{equation} \label{linratio}
	\frac{I(t)}{I(t)+A(t)} \stackrel{t \rightarrow \infty}{\longrightarrow} \rho \;,
	\end{equation}
	with
	\begin{equation}  \label{defrho}
	\rho = 1- \frac{\beta_0(1-\xi)}{\lambda_+ + \gamma_a+ \beta_0(1-\xi)(1-\mu)}\;.
	\end{equation}
	In the above equation $\lambda+$ and $\beta_0$ are related through Eq.~(\ref{betalambda}). For the parameter values of Fig. \ref{ratiograph} we have $\rho= 0.610411$, which is approximately the height of the plateau in that graph.
	
	This agreement illustrates the more general fact that the exact solution of Eq.~(\ref{linasireqs}) is a good approximation to the true solution of Eq.~ (\ref{asireqs}) at the beginning of the epidemic. Of course, the exact solution of Eqs.~(\ref{linasireqs}) breaks down as an approximation for larger times. This is also shown in Fig.~\ref{ratiograph}, because $\frac{I(t)}{I(t)+A(t)}$ is 
	not always close to the value defined in Eq.~(\ref{defrho}).

	Another consequence of Fig. \ref{ratiograph} is that the ratio of symptomatic to total infected individuals is a dynamic quantity. It cannot be included as a parameter in the model's equations as in \cite{Li}. Moreover, as shown by Eq.~(\ref{defrho}), not even for the small interval of time in which the ratio $\frac{I(t)}{I(t)+A(t)}$ is approximately constant, it equals parameter $\xi$. 
	
	We will call \textit{linear regime} the time interval in which the true solution of (\ref{asireqs}) is well approximated by the solution of (\ref{linasireqs}). For the parameter values in Figs. \ref{asirbasic} and \ref{ratiograph} the linear regime lasts approximately up to time $t=25$, in which the fractions of symptomatic and MSA individuals are already quite high.
	
	In the linear regime, any of the quantities $I(t)$, $A(t)$, $R_s(t)$ and $R(t)$ are approximated by exact solutions having the form $c_1 e^{\lambda_+ t}+ c_2 e^{\lambda_- t}$, where $c_1$ and $c_2$ are constants depending of which quantity we are calculating. As the term $e^{\lambda_- t}$ quickly tends to 0, we see that $I(t)$, $A(t)$, $R_s(t)$ and $R(t)$ are all approximated by $c_1 e^{\lambda_+ t}$, i.e., all of them are exponentially growing in the linear regime whenever $\lambda_+>0$. Most importantly, all of them grow at the same rate determined by the largest eigenvalue $\lambda_+$. For this reason, $\lambda_+$ is called the \textit{Malthusian parameter} of the model \cite{diekmann}.
	
	\subsection{Conditions for the fraction of infected individuals to decrease}
	It can be shown \cite{diekmann} that the total fraction of infected individuals $I(t)+A(t)$ in the solutions of (\ref{asireqs}) will decrease for all $t>0$ if and only if $\lambda_+ \leq 0$. The same condition is generally written in terms not of $\lambda_+$, but of the basic reproduction ratio ${\cal R}_0$. ${\cal R}_0$ is defined as the mean number of individuals infected by a single infected individual during its whole infective period if the population is entirely susceptible. If ${\cal R}_0 \leq 1$, it can be shown that the number of infected individuals will initially decrease and always decrease in the A-SIR model. It is straightforward to calculate ${\cal R}_0$ for the A-SIR model following any of the recipes given in \cite{diekmann}. The result is
	\begin{equation}
	\label{R0}{\cal R}_0 \,=\, \beta_0 \, \left[\frac{\xi}{\gamma_s} \,+\, \frac{\mu(1-\xi)}{\gamma_a}\right]\;.
	\end{equation}
	As commented before, the Malthusian parameter $\lambda_+$ will be positive if $\beta_0$ is sufficiently large. The exact condition is exactly that the right-hand side in the above equation is larger than 1, i.e. $\beta_0> \gamma_s \gamma_a/(\gamma_a \xi+\gamma_s \mu (1-\xi))$.
	
	If ${\cal R}_0>1$ and the whole population is susceptible, the total number of infected individuals will initially increase, but as the number of susceptible individual decreases, contagion becomes more difficult, and, consequently, the total number of infected will reach a maximum at some time $t_*$. In the simpler SIR model, it can be shown that $t_*$ is the time such that $S(t_*) =1/{\cal R}_0$. In the A-SIR model, no such simple condition exists, as we have two types of infected individuals and the number of one type may increase at the same time the other decreases. An instance of that is shown in Fig.~\ref{asirbasic} in the interval between the maximum point of $I$ and the maximum point of $A$. 
	
	Gaeta \cite{Gaeta} provided conditions for each of the fractions $I$ and $A$ to decrease. As Fig. \ref{ratiograph} illustrates, whenever $\gamma_s> \gamma_a$ and $t$ is large enough, the fraction of MSA individuals is much larger than the fraction of symptomatic infected individuals, so that $I(t)+A(t) \approx A(t)$. We may use this fact to give an \textit{approximate} condition for the fraction of total infected individuals to decrease.
	
	In fact, the third equation in (\ref{asireqs}) shows that $A(t)$ decreases whenever
	\[S(t)< \frac{\gamma_a A}{\beta_0(1-\xi)(I+\mu A)}\,=\, \frac{\gamma_a}{\beta_0(1-\xi) \left[ \frac{I+A}{A}- (1-\mu)\right]}\;.\]
	Substituting $\frac{I+A}{A}$ in the above formula for its approximate value 1 for large times, we get the \textit{approximate} condition
	\begin{equation}
	\label{approxcond}S(t) \,<\, \frac{\gamma_a}{\beta_0(1-\xi)\mu}
	\end{equation}
	for the decrease of the total number of infected individuals. We stress that the above condition is approximate and holds whenever $\gamma_s>\gamma_a$ and $t$ is large enough.
	
	Notice that the \textit{threshold} in the right-hand side of (\ref{approxcond}) is higher for smaller $\beta_0$. As $S(t)$ starts close to 1 and decreases, the threshold will be easier to attain if $\beta_0$ is smaller. In other words, if $\beta_0$ is small, less people have to be infected in order that the number of infected individuals starts to decrease.
	
	\subsection{The asymptotic equilibrium}
	All solutions of the A-SIR equations (\ref{asireqs}) converge as $t \rightarrow \infty$ to the disease-free equilibrium in which $S=S_{\infty}$, $I$ and $A$ are both null and $R=1-S_{\infty}$. For a very contagious virus such as SARS-Cov-2, i.e. for large $\beta_0$, $S_{\infty}$ is close to 0. In other words, almost the entire population is eventually infected in an unmitigated epidemic caused by a sufficiently  contagious virus. This is illustrated in Fig. \ref{asirbasic}, in which $S_{\infty}$ can be numerically calculated to be $0.0181302$.
	
	Individuals exit the susceptible compartment of the model 
	either as symptomatic or MSA infected, and they do so at ratios $\xi$ and $1-\xi$, respectively. Since they will eventually become either symptomatic removed, or MSA removed, 
	then the symptomatic removed and MSA removed fractions at equilibrium obey
	\[\frac{R_s(\infty)}{R_a(\infty)} \,=\, \frac{\xi}{1-\xi}\;.\] 
	Moreover, $R_s(\infty)+R_a(\infty)=1-S_{\infty}$. Solving the set formed by the latter two equations, we obtain that  $R_s(\infty)=(1-S_{\infty}) \xi$. In the important case of a very contagious virus,
	\begin{equation}
	\label{asympRs}
	R_s(\infty) \approx  \xi \;.
	\end{equation}
	This approximation is also well illustrated in Fig. \ref{asirbasic}.
	
	\section{Fitting the A-SIR model to real COVID-19 epidemic data}\label{secfit}
	According to Crisanti \cite{crisanti} and Fenga \cite{fenga}, the lack of efficient testing has been responsible for a substantial underestimation of the number of cases in the COVID-19 epidemic in Italy. In particular, probably due to testing preferentially the most severe cases, the mortality rate in the regions of Lombardy and Emilia-Romagna was three times larger than in the neighboring region of Veneto, in which testing for COVID-19 was widespread \cite{crisanti}. The same problem caused by lack of tests is reported also in Brazil and elsewhere.
	
\begin{figure}
		\begin{center}
			\includegraphics[width= \textwidth]{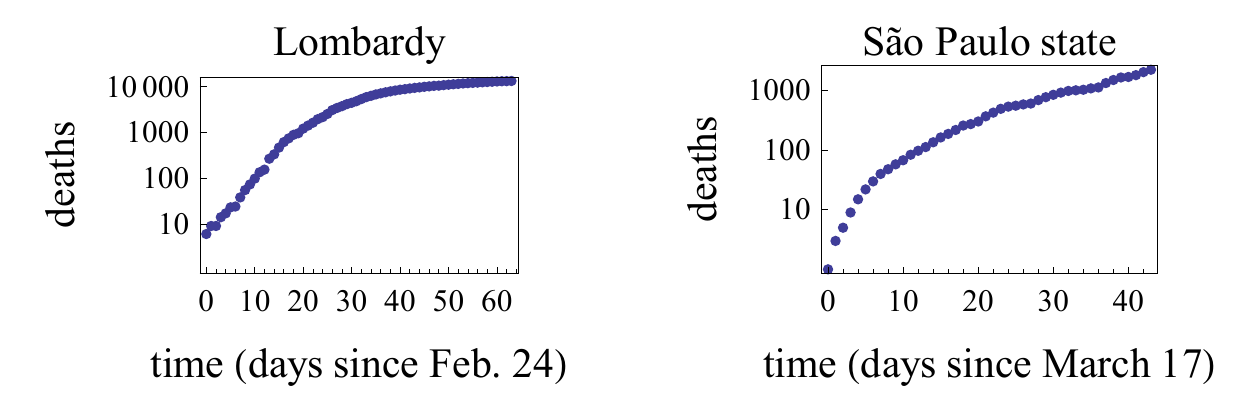}
			\caption{\label{figdeaths} The number of accumulated deaths due to COVID-19 in Lombardy since Feb. 24, 2020 (left) and in S\~ao Paulo state since Mar. 17, 2020 (right). Data collected respectively in \cite{datiitalia} and \cite{dadossp}.}
		\end{center}
	\end{figure}

	Because of this, we believe that the cumulative number of deaths is a much more faithful indicator of the evolution of the epidemics than the number of confirmed cases, both in Lombardy and in S\~ao Paulo. We chose the deaths data in both locations as the sources for our fitting. This approach was also taken e.g. in \cite{imperial13}. The official deaths data, plotted in Fig.~\ref{figdeaths}, were collected respectively in \cite{datiitalia} and \cite{dadossp}, along with the official numbers of confirmed cases.
	
	We need one extra parameter, 
	$\omega$,
	to relate the outcome of the model to the number of reported deaths used here.  In fact, the A-SIR model does not make any prediction for the population fraction $D(t)$ of individuals dead due to COVID-19 up to time $t$. As only the symptomatic cases may die, it is natural to suppose that
	\begin{equation}
	\label{DRs}
	D(t) \,=\, \omega R_s(t) \;,
	\end{equation} 
	where $\omega \in (0,1)$ is thus interpreted as the \textit{case fatality rate}. The value for $\omega$ must also be found. As the lack of tests is a reality, an examination of data for several countries, as in \cite{worldmeters}, shows that the ratio of deaths to confirmed cases varies broadly
	among them. 
	
	Before entering into details, we describe the fitting procedure in general. Both in Lombardy and in S\~ao Paulo state, the epidemic started uncontrolled. Noticing the logarithmic vertical scale, we can see in both panels of Fig.~\ref{figdeaths} that in the first days the number of deaths seemed to increase exponentially, as expected for the linear regime, see subsection \ref{sublinear}. We will call these first days as the phase of \textit{uncontrolled epidemic}. After this phase, in both locations the number of deaths started to increase at a lower rate. One of the important things will be to assess whether this lower rate is a natural consequence of the A-SIR model, or if it is due to the mitigation measures adopted in both locations.
	
	As will be fully explained in the following, the first step will be using the deaths data in the uncontrolled epidemic phase to estimate the values of $\lambda_+$ and $\omega$. 
	
	In the next steps, for each phase of social distancing we will have another parameter $\epsilon$ indicating the intensity of the adopted measures. In Lombardy we will consider two phases of different intensities $\epsilon_1$ and $\epsilon_2$ for the period of social distancing. In S\~ao Paulo state, only one phase of social distancing will be considered. We will see that it is possible to use the number of deaths data to obtain estimates for $\xi$ and for the intensities $\epsilon_i$. Although in general an optimal choice for these parameters exists, we will see that many possible choices are almost as good for the purpose of fitting the data with the model. It results that more than one possible good fit of the model to the data exists. We will explore the consequences of this approximate degeneracy in the optimization procedure.
	
\begin{figure}
		\begin{center}
			\includegraphics[width= \textwidth]{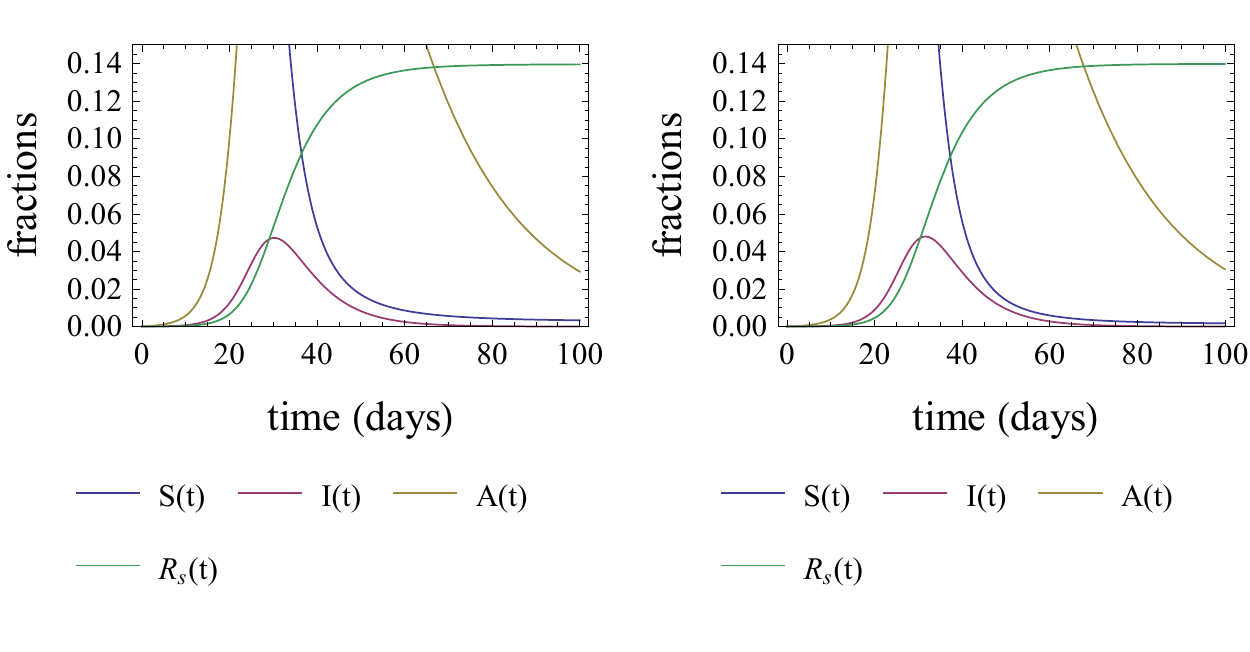}
			\caption{\label{figindist} Two almost indistinguishable solutions of the A-SIR model with $\gamma_s=1/7$, $\gamma_a=1/21$, $\xi=0.16$, $\lambda_+= 0.3$. The left panel was produced with $\mu=0.25$ and the right panel with $\mu=0.6$. Observe that $\beta_0$ is calculated by Eq.~(\ref{betalambda}), thus it gets disparate values in the two simulations;		1.06995 for the solution in the left panel, and 0.555397 for the
				solution in the right panel. The initial conditions were $S(0)=1$, $I(0)= A(0)= 6.22 \times 10^{-6}$ and $R_s(0)=0$ for both cases.}
		\end{center}
	\end{figure}

	Since parameters $\gamma_a$ and $\gamma_e$ are fixed, as already explained, and $\beta_0$ will be 
	related to $\lambda_+$ by
	Eq.~(\ref{betalambda}),
	the
	parameter $\mu$ still remains undetermined. After several
	experiences we noticed that,  
	\textit{as long as $\lambda_+$ is determined and $\beta_0$ is related to it by (\ref{betalambda})}, 
	the results of the model are 
	to large extent independent of $\mu$. This is a consequence of the fact that the initial behavior of the deaths number 
	according to the model is dictated by the linear regime, i.e., $\lambda_+$, whereas 
	the final behavior is dictated by $\xi$, see Eq.~(\ref{asympRs}). The above phenomenon is illustrated in Fig.~\ref{figindist},
	in which the results of A-SIR numerical solutions with two different values of $\mu$ show no noticeable differences. Thus, there 
	will be no
	problem in adopting for $\mu$ any fixed value. We assume
	$\mu=0.5$ for the rest of this paper. 
	
	The goodness of a fit between the number of deaths reported in the data and the A-SIR model
	will be quantified by a \textit{cost function} to be minimized. For future reference, 
	our choice of cost function is
	\begin{equation}  \label{cost}
	c_d(\lambda_+, \xi, \mu, \omega) \,=\, \frac{N}{d} \, \sqrt{\sum_{i=0}^{d} (D(i)- \omega R_s(i))^2}\;.
	\end{equation}
	In the above formula, besides other notations already introduced, $N$ is the total population of the location (Lomabardy or S\~ao Paulo state) and $d$ is the last day up to which the numbers of reported deaths $D(i)$ are being compared to the prediction of the model $\omega R_s(i)$. We do not indicate the dependence of the cost function on the fixed parameters $\gamma_a=1/21$ and $\gamma_s=1/7$. The cost function also depends on the initial conditions to be used in the numerical solution of the model's differential equations. The initial conditions will be specified when needed. The value of $d$ depends on which data is used for estimating the parameters. For example, for estimating $\lambda_+$ and $\omega$ we will take $d$ to be the last day of the uncontrolled epidemic phase. For estimating the intensity of social distancing measures, we will use a large value for $d$. Of course neither the square root, nor the multiplication by $N/d$ are necessary in the definition of the cost function, but they are included for convenience.
	
	We will describe first the fitting procedure for Lombardy, where more data is available and then proceed to S\~ao Paulo state.
	
	\subsection{Fitting the epidemic in Lombardy}
	
	The national lockdown in Italy started on March 9, 2020, day number 14 after Feb. 24, the start date of the Italian time-series. We will define Mar. 9, 2020 to be the end of the uncontrolled epidemic phase for Lombardy. 
	
	For the numerical solution of Eqs.~(\ref{asireqs}) in Lombardy we use initial conditions 
	\begin{align}
	S(0)&=1,  \;\;\; I(0)= 1.66 \times 10^{-5}, \nonumber\\
	A(0)&= \frac{1- \rho}{\rho} I(0), \;\;\; R_s(0)=\frac{1}{\omega} \, 6 \times 10^{-7}\;. \label{lombinit}
	\end{align}
	Here we used that the population of Lombardy is $10^7$ inhabitants and that the number of confirmed COVID-19 cases at day 0 was 166 \cite{datiitalia}, accounting for the value of $I(0)$. As the initial number of MSA is unknown, our choice for $A(0)$ is a natural one such that the fraction $I(t)/(I(t)+A(t))$ is equal to $\rho$ already at day 0, see (\ref{linratio}) and (\ref{defrho}). The initial condition for $R_s(0)$ is also the natural one using the number of people dead due to COVID-19 at day 0, obtained from the data \cite{datiitalia}, and taking into account Eq.~(\ref{DRs}).
	
	We stress that the initial conditions for $I(0)$ and $A(0)$ are the only places in our fitting procedure where some information on the number of confirmed cases is used. We will not use such information in any other day. It is conceivable that at day 0 the number of confirmed cases is more reliable than in later days. Moreover, the exact values of $I(0)$ and $A(0)$ do not matter so much, as $I(t)$ and $A(t)$ initially grow exponentially by a rate to be determined by the number of deaths.
	
\begin{figure}
		\begin{center}
			\includegraphics[width= 0.7 \textwidth]{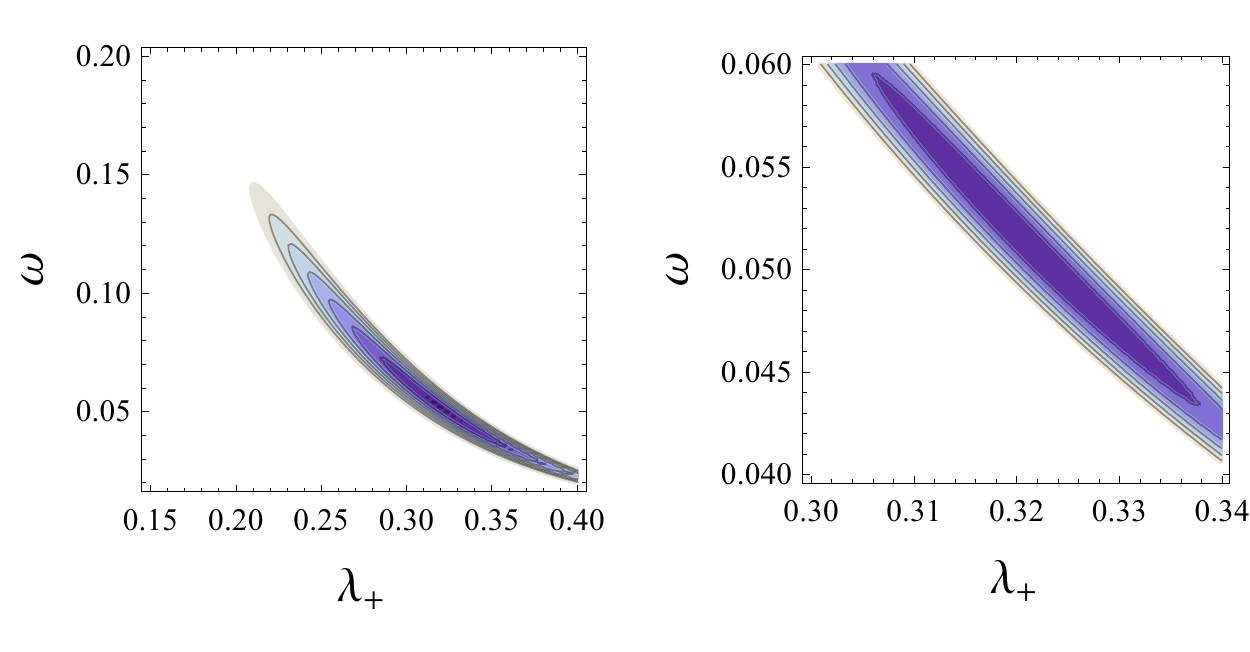}
			\caption{\label{detlambom1} Left:
				Contour plots of $c_{14}(\lambda_+, 0.1, 0.5, \omega)$ 
				for a broad range in $\lambda_+$ and $\omega$.	Right:  Zoom-up to the region of interest (darkest area of the left panel).	Darker colors are smaller values and lighter colors are higher values of the cost function. In the 
				white area the cost function
				has larger values than in the colored area. It seems clear that the function has a global minimum in the darker area of the right panel. The numerically determined location of the minimum point is $\lambda_+= 0.3216$, $\omega=0.051$.}
		\end{center}
	\end{figure}

	In Fig. \ref{detlambom1} we show contour plots of the objective function $c_{14}(\lambda_+, 0.1, 0.5, \omega)$. Observe that we are considering only the deaths data up to day 14, i.e. in the uncontrolled epidemic phase. We arbitrarily fixed $\mu=0.5$ and $\xi=0.1$. The plots show that, for the chosen values of $\mu$ and $\xi$, in a large region in the $(\lambda_+, \omega)$ plane the cost function has a single local minimum, which is probably a global minimum. The set of parameter values that minimizes the cost function, i.e., provide the best fit of the model to the data up to day 14, is $\lambda_+= 0.3216$, $\omega=0.051$.
	
\begin{figure}
		\begin{center}
			\includegraphics[width= 0.7 \textwidth]{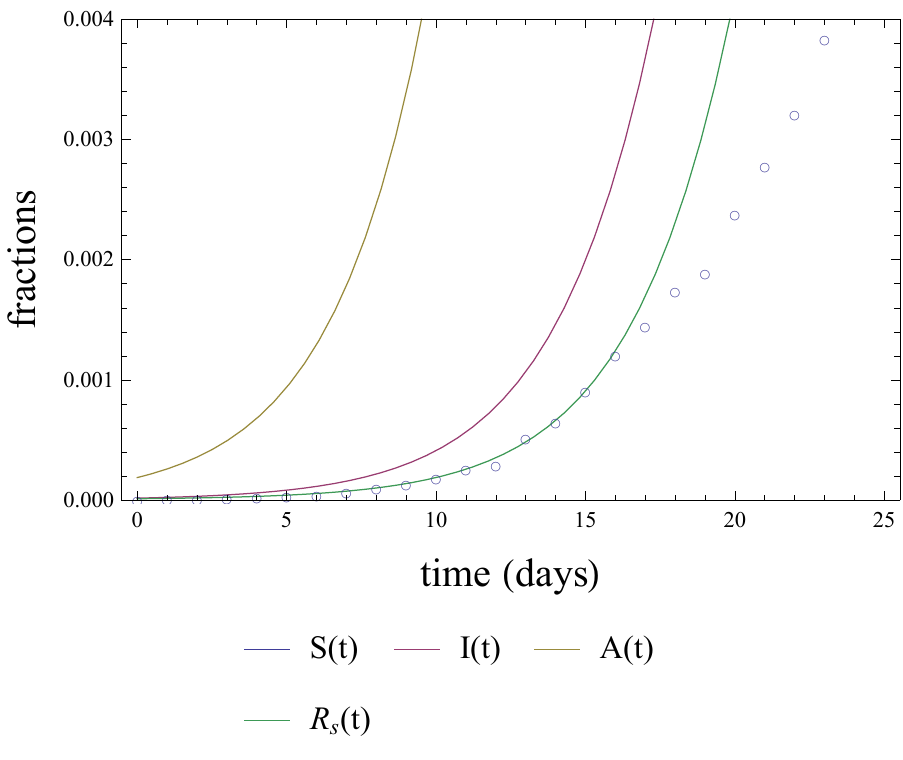}
			\caption{\label{uncontrol1} Results of the A-SIR model, 
				Eqs.~ (\ref{asireqs}) with initial conditions (\ref{lombinit}) and parameters $\gamma_s=1/7$, $\gamma_a=1/21$, $\mu=0.5$, $\xi=0.1$, $\lambda_+=0.3216$, $\omega= 0.051$. The latter two were chosen 
				to minimize the cost function, $c_{14}$, using the former four parameters. The blue dots correspond to the number of deaths reported in pandemic data divided
				by $\omega$. 
				The graph of $S(t)$ lies out of the range shown in the figure.}
		\end{center}
	\end{figure}

	Fig.~\ref{uncontrol1} compares the epidemic data with the solution of Eqs.~(\ref{asireqs}) with initial conditions (\ref{lombinit}) and parameters $\gamma_s=1/7$, $\gamma_a=1/21$ (fixed), $\mu=0.5$, $\xi=0.1$ (arbitrarily chosen) and $\lambda_+=0.3216$, $\omega= 0.051$ (optimally chosen with respect to the preceding values). The dots in the figure are the data for $D(i)$ divided by $\omega$, and should thus approximate the curve for $R_s(t)$ up to $t=14$. Numerical experiments (not presented for conciseness) with different values for $\mu$ and $\xi$ 
	confirm that the graph $R_s(t)$ up to $t=14$, to be adjusted to data, almost does not change. Thus, the fit shown in the figure remains equally good independent of the values of
	$\mu$ and $\xi$.  As already mentioned $\mu$ is quite irrelevant as far as $\beta_0$ is calculated as a function of $\lambda_+$. Also, the value of $\xi$ is not relevant for reproducing
	the first days of the epidemic. 
	
	It is well clear by Fig. \ref{uncontrol1} that after day 14 (start of the national lockdown in Italy) the data 
	increase less than the predicted $R_s$ based on an uncontrolled epidemic. This is a good evidence that the lockdown was effectively important in reducing the number of deaths due to COVID-19 in Lombardy.
	
\begin{figure}
		\begin{center}
			\includegraphics[width= 0.5 \textwidth]{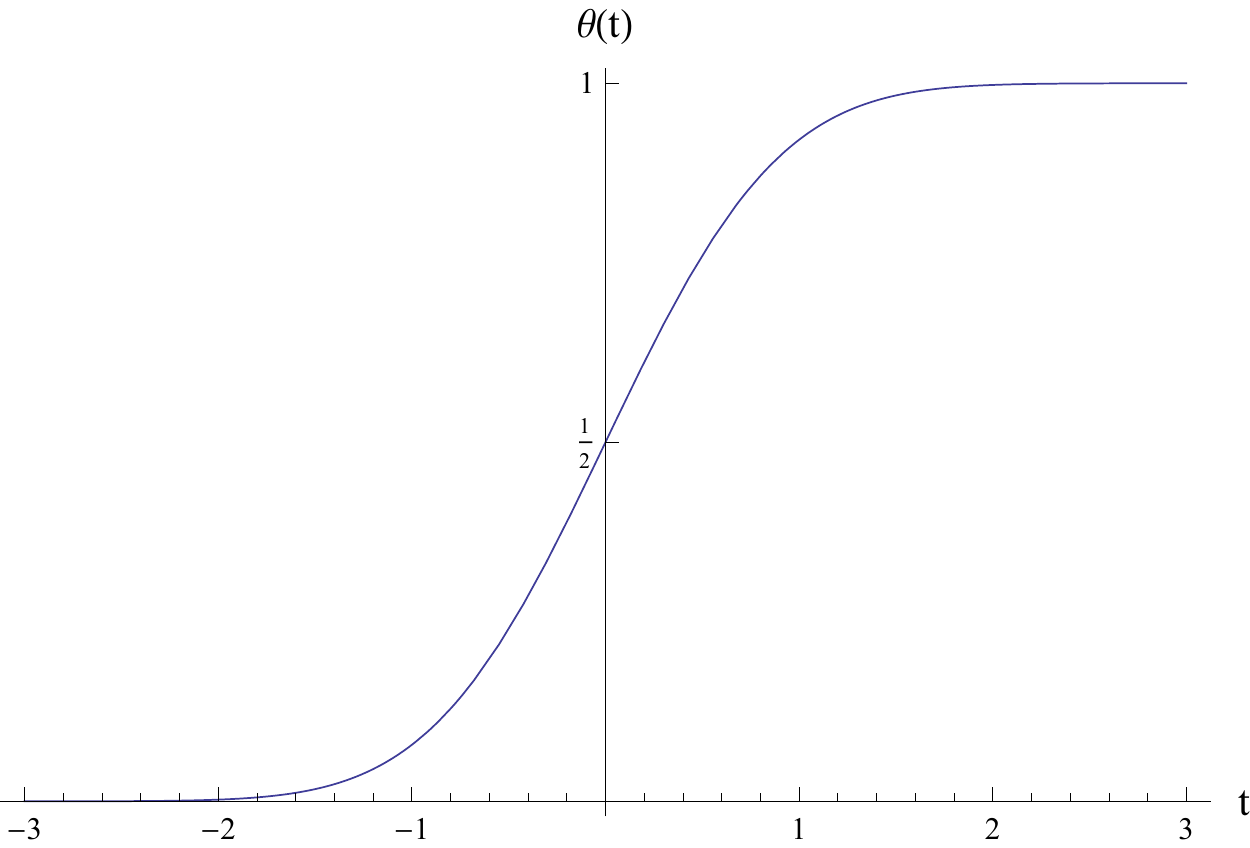}
			\caption{\label{figtheta} Graph of the transition function $\theta(t)$ considered  in Eq.~(\ref{deftheta}). }
		\end{center}
	\end{figure}

	We will incorporate the effect of social distancing measures in our model by introducing a decrease of the infection rate $\beta_0$ to a smaller value $\epsilon_1 \beta_0$, where $0<\epsilon_1 <1$. More precisely, in order to avoid introducing a discontinuous function into the system of differential equations, we replace $\beta_0$ in equations (\ref{asireqs}) by the smooth function
	\begin{equation}
	\beta(t) \,=\, \beta_0 r(t)
	\end{equation}
	with
	\begin{equation}
	r(t)=1 - (1 - \epsilon_1) \theta(t-14)\;.
	\end{equation}
	In the above formula, $\theta(t)$ may be any continuous approximation of the unit step function. We used
	\begin{equation} \label{deftheta}
	\theta(t)\,=\, \frac{1}{2}\, (1+ \mathrm{erf}\, (t)) \;,
	\end{equation}
	where $\textrm{erf}\,(z)=2/\sqrt{\pi} \int_0^z e^{-t^2} \, dt$ is the integral of the normal distribution. The graph of $\theta(t)$ is shown in Fig. \ref{figtheta}. Any similar continuous function switching from values close to 0 to values close to 1 in an interval around 0 of size 1 can be used without relevant changes. The important thing is that $\beta(t) \approx \beta_0$ for $t<14$ and $\beta(t) \approx \epsilon_1 \beta_0$ for $t>14$.
	
	In the second step of our fitting procedure, we will 
	fix the values of $\lambda_+$ and $\omega$ already 
	estimated and evaluate
	the values for the intensity $\epsilon_1$ of the first phase of the lockdown, and parameter $\xi$, which was not relevant in the uncontrolled epidemic phase. 
	We referred above to the first phase of the lockdown, because a strengthening of 
	it occurred on Mar. 22, day 27 after Feb. 24. Therefore, to estimate
	$\epsilon_1$ and $\xi$ we will 
	use $d=27$ in the cost function, Eq.~(\ref{cost}). 
	
\begin{figure}
		\begin{center}
			\includegraphics[width= \textwidth]{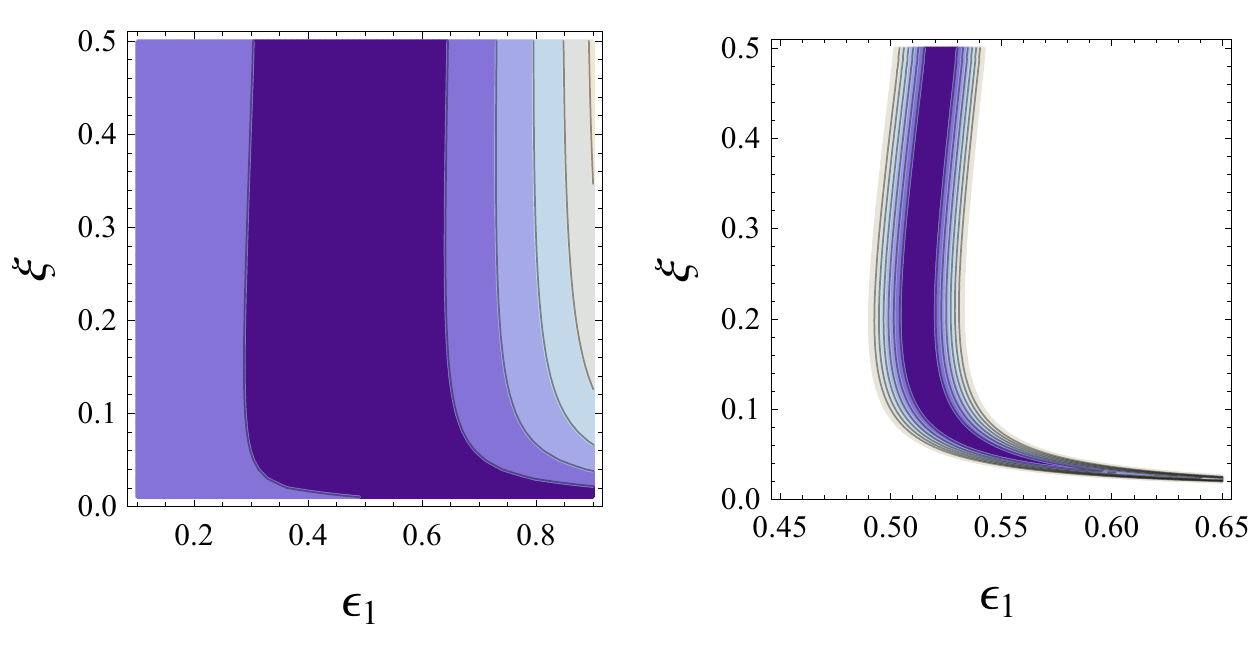}
			\caption{\label{lombquarant1} 	
				Left: Contour plot 
				of $c_{27}$ as a function of $\epsilon_1$ and $\xi$ in a 
				large region of parameter choices.
				Right: Zoom to the
				smaller region contained in the dark area of the left panel plot. The remaining parameters have the same values as 
				in Fig.~\ref{uncontrol1}. Darker colors are smaller values and lighter colors 
				are higher values of the cost function. The global minimum is located at $(0.513,0.256)$, 
				but the cost function is rather close to the minimum in the large darker region in the right panel.}
		\end{center}
	\end{figure}

	Fig.~\ref{lombquarant1} shows contour plots of the cost function $c_{27}$ with parameters $\gamma_s$, $\gamma_a$, $\mu$,  $\lambda_+$ and $\omega$ having the same values as in Fig.~\ref{uncontrol1}, but allowing now variation of $\xi$ and of the new parameter $\epsilon_1$. The left panel shows a larger region in $(\epsilon_1, \xi)$ plane and the right panel shows in detail the region in which $c_{27}$ attains its smallest values. We notice that, contrary to the analogous plots in Fig.~\ref{detlambom1}, in which a clear 
	global minimum is evidenced, this time
	the region where
	the cost function is close to minimum seems like a large "canyon".
	
	Referring to the darker area in the right panel of Fig.~\ref{lombquarant1}, we will call \textit{optimistic} choices for $(\epsilon_1,\xi)$ those points in the darker region with smaller values of $\xi$. In fact, because of Eqs.~(\ref{asympRs}) and (\ref{DRs}), such a choice will lead to a small expected number of dead people at the end of the epidemic. On the contrary, points in the darker region with larger values of $\xi$ will be called \textit{pessimistic choices} for $(\epsilon_1,\xi)$. 
	
\begin{figure}
		\begin{center}
			\includegraphics[width= \textwidth]{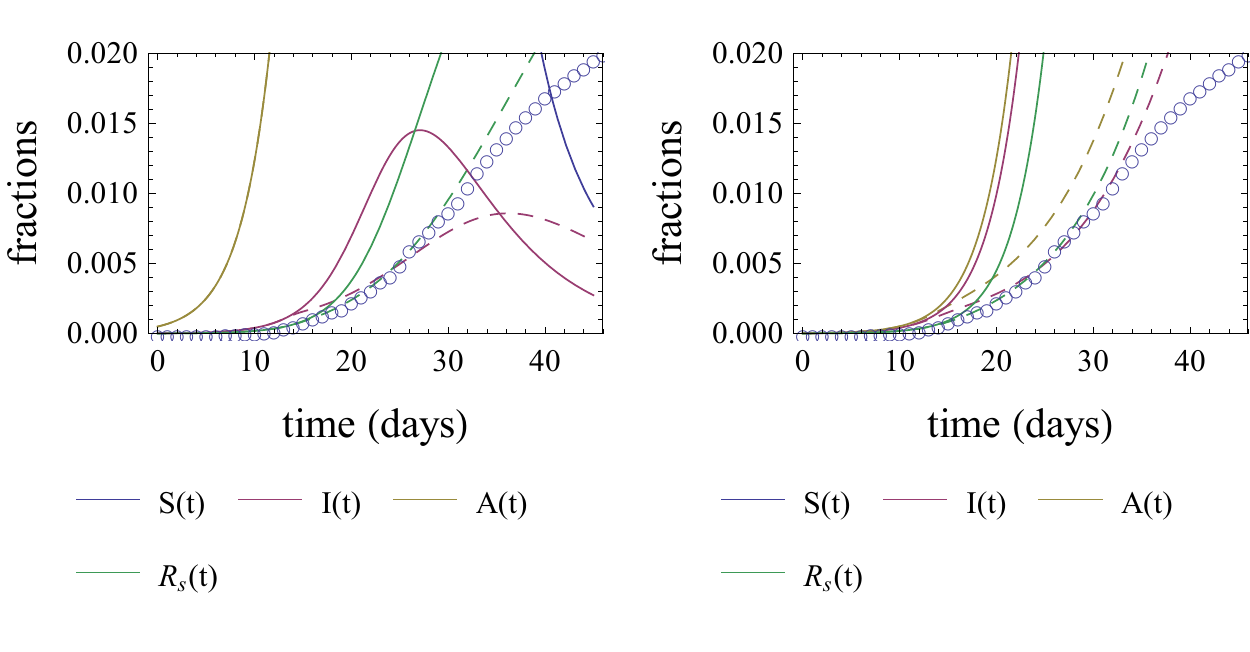}
			\caption{\label{graflombq1} Plots of two solutions to Eqs.~(\ref{asireqs}) with different possibilities for the effect of the social distancing measures introduced by the lockdown of Mar. 9. In both panels, the full lines show, for comparison, what would be the solutions if the epidemic remained uncontrolled, whereas the dashed lines in corresponding colors show the effect of social distancing measures beginning at $t=14$. The results on the
				left panel were
				produced with 
				$\epsilon_1=0.566$, $\xi= 0.04$, and those on the right panel	were obtained 
				with $\epsilon_1=0.52$, $\xi=0.5$. The remaining parameters are the optimal ones 
				determined using the results presented in
				Fig.~\ref{uncontrol1}. The choice for the left panel is an example of what we 
				called an optimistic choice. The right panel corresponds to a pessimistic choice for 
				the parameters. }
		\end{center}
	\end{figure}

	We show in Fig.~\ref{graflombq1} plots of two solutions to Eqs.~(\ref{asireqs}) with optimization
	up to day 27.  Both choices for $(\epsilon_1,\xi)$ are almost equally good in fitting the data for days 0 to 27, however the results on the left panel were obtained with an optimistic choice for $(\epsilon_1,\xi)$, while the results on
	the right panel were obtained by considering  a
	pessimistic choice.
	
	In both panels of Fig. \ref{lombquarant1} we also see that after day 27 the green dashed curve of $R_s(t)$ grows faster than the data points. This suggests us that the strengthening of the lockdown in Lombardy after Mar. 22 did produce effects in slowing down the number of deaths.
	
	We will then introduce a further parameter $\epsilon_2$ such that $\beta(t)= \epsilon_2 \beta_0$ for $t>27$. 
	In order to see the effects of relaxing the social isolation measures, we will also restore $\beta(t)$ smoothly to its value $\beta_0$ for 
	$t>100$. Precisely, we will take
	\begin{equation}
	r(t)=1 - (1 - \epsilon_1) \theta(t-14)- (\epsilon_1-\epsilon_2) \theta(t-27)+ (1-\epsilon_2)\theta(t-100)\;.
	\end{equation}
	
\begin{figure} 
		\begin{center}
			\includegraphics[width= \textwidth]{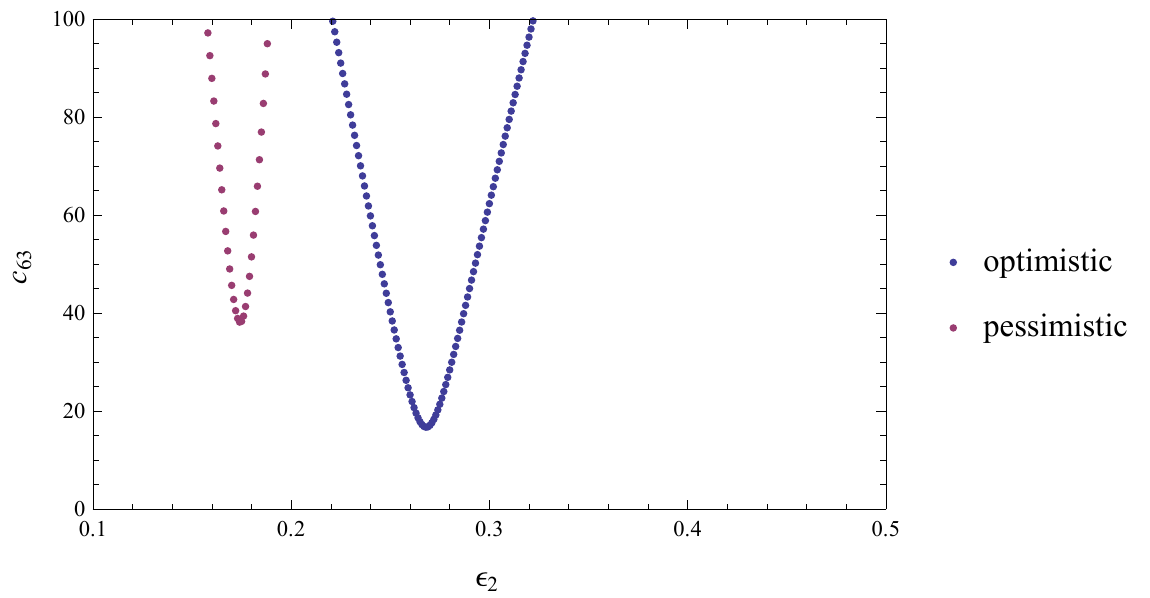}
			\caption{\label{deteps2} Plots of the cost function $c_{63}$ for several values of $\epsilon_2$ and two choices of $(\epsilon_1,\xi)$, the same optimistic	(blue)	and pessimistic (red) choices already used in the panels of Fig.~\ref{graflombq1}. The remaining parameter values are 	the ones used in Fig.~\ref{uncontrol1}. The minima of the cost function occur at 
				$\epsilon_2=0.268$ (optimistic case), and $\epsilon_2=0.186$ (pessimistic case).}
		\end{center}
	\end{figure}

	The value for $\epsilon_2$ will be determined by minimizing the cost function up to day 63, April 27. It turns out that the best choice for $\epsilon_2$ will depend on the choice made for $\xi$ and $\epsilon_1$. In Fig. \ref{deteps2} we show the behavior of
	the cost function, $c_{63}$, when varying $\epsilon_2$ for both the optimistic and pessimistic choices for $(\epsilon_1,\xi)$ already considered in Fig. \ref{graflombq1}.
	
\begin{figure} 
		\begin{center}
			\includegraphics[width= 0.9\textwidth]{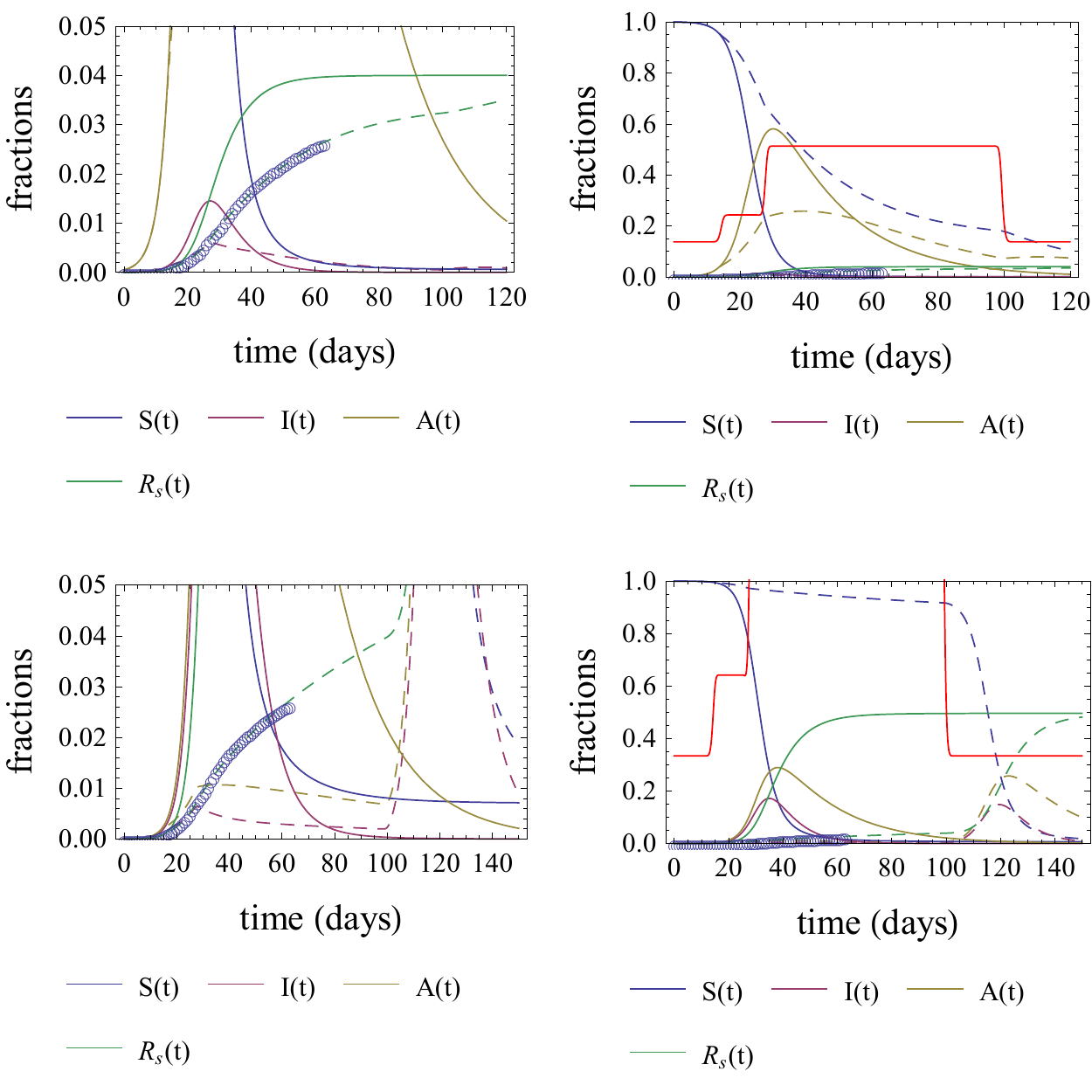}
			\caption{\label{finallomb} Expected outcomes for the COVID-19 epidemic in Lombardy for up to 
				150 days after Feb. 24, compared with the deaths data. 
				We are supposing, as a possibility, that the social isolation measures are completely relaxed at day 100, June 3.
				An optimistic outcome	is presented
				in the first row, and a pessimistic	in the second row. The plots in the left and right columns differ only
				in the range of the vertical scale. The parameter values valid for both rows are: $\gamma_s=1/7$, $\gamma_a=1/21$, $\lambda_+= 0.3216$, $\omega=0.051$, $\mu=0.5$, see Fig. \ref{uncontrol1}. For the first row only, 
				$\epsilon_1=0.566$, $\xi=0.04$, $\epsilon_2=0.268$. For the second row only, $\epsilon_1=0.52$, $\xi=0.5$, $\epsilon_2=0.186$. The red lines appearing in the second column are the graphs of $\frac{\gamma_a}{\beta(t) (1-\xi) \mu}$, see Eq.~(\ref{approxcond}). Notice that with a good approximation, when the graph of $S(t)$ is below 
				the red line, the fraction of MSA infected individuals decreases.}
		\end{center}
	\end{figure}

	As the final result of the fitting procedure for Lombardy, we show in Fig. \ref{finallomb} the graphs of the solutions to the A-SIR model with two different sets of parameter values such that the model fits rather well the deaths data. 
	
	A conclusion to be drawn from the results in Fig. \ref{finallomb} is that the A-SIR model can fit the deaths data for the epidemic in Lombardy extremely well if we consider that the social distancing measures adopted there were less intense from Mar. 9 to Mar. 22, days 14 to 27, and more intense from Mar. 22 to 
	June 3, days 27 to 100. On the other hand, the goodness of the fits does not tell us very much about the future, when the social distancing measures are relaxed. Besides the two parameter choices shown, we have many others almost as good as them. The choices shown here are extreme in the sense that 
	$\xi=0.04$ 
	is not too much above the minimum value of $\xi$ for the points in the darker region in the right panel of Fig. \ref{lombquarant1}, and $\xi=0.5$ is the maximum value in the same region. Due to Eqs.~(\ref{DRs}) and (\ref{asympRs}), these are also 
	close to the extreme possibilities for the number of deaths at the end of the epidemics.
	
	The optimistic choice, corresponding to the first row of Fig. \ref{finallomb}, is such that after complete relaxation of the social distancing, the number of cases continues decreasing and the number of accumulated deaths increases slowly. This happens because the fraction of susceptible individuals 
	since day 38 falls below the threshold in the right-hand side of Eq.~(\ref{approxcond}) and does not increase very much above it after the social isolation measures are removed at day 100.

	On the other hand, for the pessimistic choice, depicted in the second row of Fig.~\ref{finallomb}, the susceptible fraction is below the threshold in the right-hand side of Eq.~(\ref{approxcond}) between days 27 and 
	100. The red line representing the threshold is not shown in the figure in this time interval, because the threshold is larger than 1.  But the susceptible fraction stays high above the threshold after day 100, when social isolation measures are removed. As a consequence, there is a fast increase of the number of cases after day 100, i.e. a second wave of COVID-19 cases, potentially almost as intense as it would have been if the epidemic remained uncontrolled since the beginning.
	
	If we had more knowledge either on the value of $\xi$ or on the values of $\epsilon_1$ and $\epsilon_2$, we might choose among the most optimistic, the most pessimistic, or some intermediate possibility between them. In the lack of such knowledge, we present all possibilities.
	
	The value of $\xi$ could be estimated either by clinic research, or a large scale population screening. We believe that such studies are on their ways, and hope that their results might help us removing the uncertainties.
	
	As far as $\epsilon_1$ and $\epsilon_2$ are concerned, some attempts have been made to measure the intensity of the social distancing using cell phone localization data, in particular for the regions in Italy \cite{mobchanges}. A problem is that we do not know how to relate these measures with the decreases $\epsilon_1$ and $\epsilon_2$ in the infection rate. 
	
	\subsection{Fitting the epidemic in S\~ao Paulo state}
	In Brazil the first imported case of COVID-19 was identified in the city of S\~ao Paulo on Feb. 26, 2020. The first official death, also in S\~ao Paulo, occurred on March 17, 2020. For this reason, we chose that date as day 0 for the epidemic in Brazil. Although there has been up to now no nation-wide social distancing measures in Brazil, many state governors and mayors, including the governor of S\~ao Paulo state, decreed such measures. It is difficult to identify a clear starting date, as measures were gradual, but it seems reasonable to choose Mar. 23, day number 6 after the first death, as the end of the uncontrolled epidemic phase in the state of S\~ao Paulo. Mar. 23 was, in fact, the first day in which all schools were closed in S\~ao Paulo state.
	
	After the start of the social distancing measures, cell phone localization data \cite{inloco} show a tendency of slowly decreasing efficiency of these measures. Although some metropolitan areas in Brazil have already declared lockdown, no such measure has been declared at S\~ao Paulo up to the time of the latest data shown in this paper, Apr. 29. As a consequence, we prefer to use a single reduction of infection rates in fitting the epidemic in S\~ao Paulo.
	
	The fitting procedure for S\~ao Paulo is thus similar to the one for Lombardy, with the exception that we use only one social distancing reduction factor $\epsilon_1$ for the entire period of the data. Taking into account that the population of the state of S\~ao Paulo is $N= 4.6 \times 10^7$ people, the initial conditions used for the numerical solution of the ODEs~(\ref{asireqs}) were
	\begin{align}
	S(0)&=1,  \;\;\; I(0)= 3.56522 \times 10^{-6}, \nonumber\\
	A(0)&= \frac{1- \rho}{\rho} I(0), \;\;\; R_s(0)=\frac{1}{\omega} 2.17391\times 10^{-8}\;. \label{spinit}
	\end{align} 
	The choice of these conditions is analogous to what was done for Lombardy.
	
\begin{figure}
		\begin{center}
			\includegraphics[width= 0.7\textwidth]{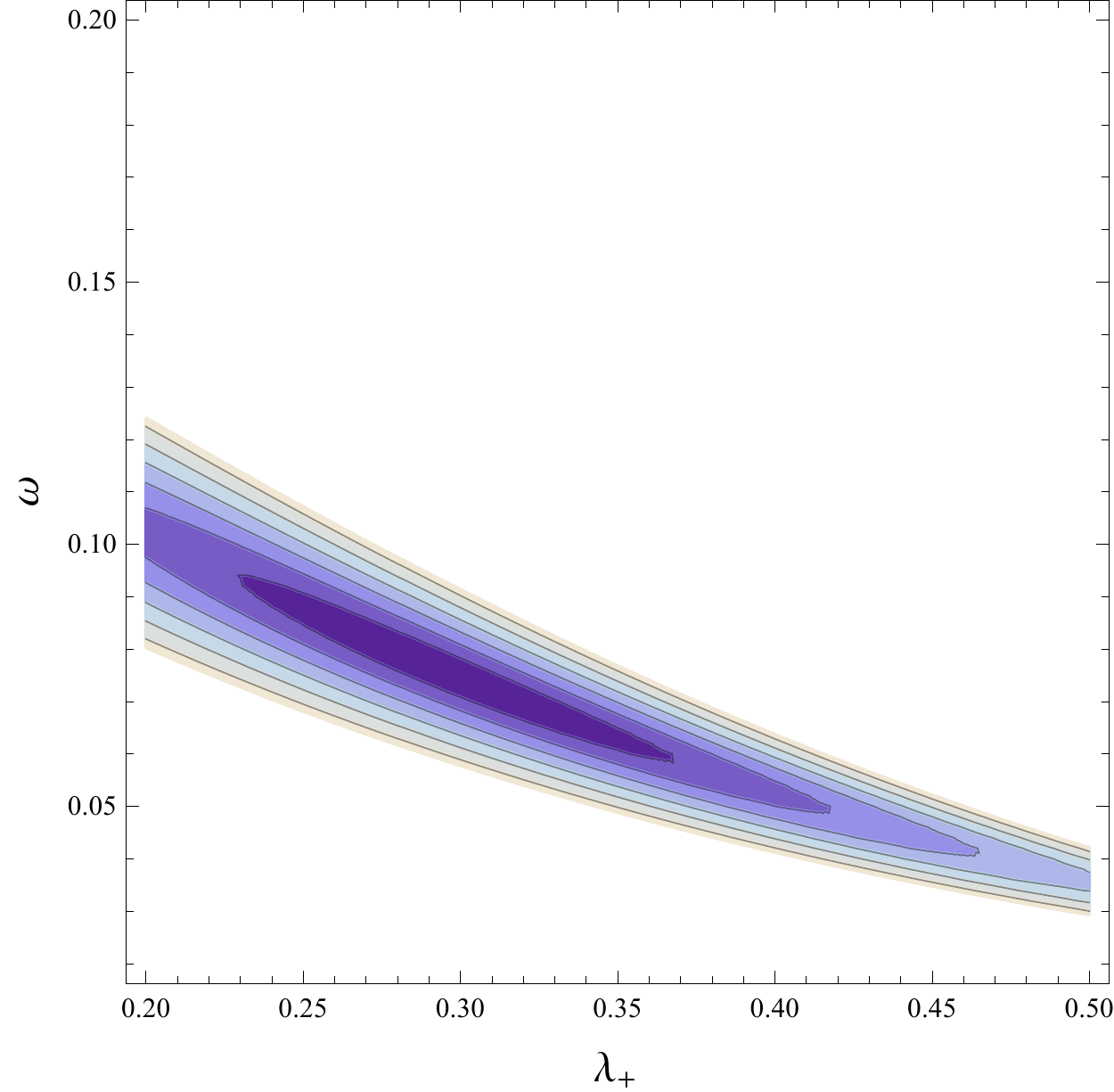}
			\caption{\label{detlambom2} Contour plot of $c_{6}(\lambda_+, 0.1, 0.5, \omega)$ for the state of S\~ao Paulo. Darker colors are smaller values and lighter colors are higher values of the cost function. In the large white area the cost fuction has larger values than in the colored area. It seems clear that the function has a global minimum in the darker area. The numerically determined location of the minimum point is $\lambda_+= 0.302$, $\omega=0.074$.}
		\end{center}
	\end{figure}

	The uncontrolled epidemic phase lasted from day 0 (Mar. 17) to day 6 (Mar. 23). We used the deaths during that phase to obtain estimates for $\lambda_+$ and $\omega$. Fig. \ref{detlambom2} shows that the cost function $c_6(\lambda_+,0.1,0.5,\omega)$ apparently has a global minimum at $(\lambda_+,\omega)= (0.302,0.074)$.
	
\begin{figure} 
		\begin{center}
			\includegraphics[width= 0.7\textwidth]{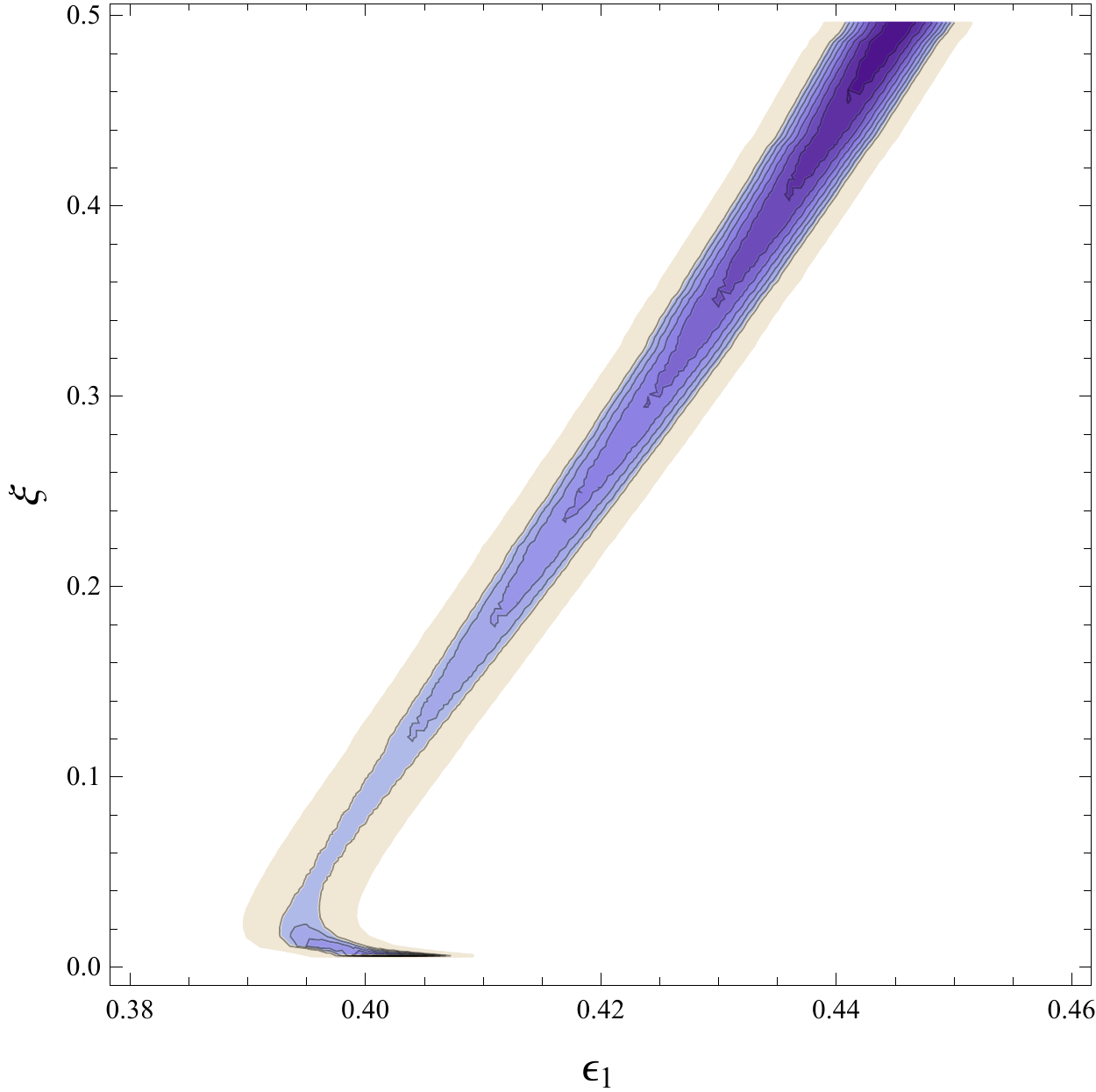}
			\caption{\label{spquarant} Contour plot of $c_{43}$ as a function of $\epsilon_1$ and $\xi$. The remaining parameters are $\gamma_s=1/7$, $\gamma_a=1/21$ (fixed), $\mu=0.5$ (arbitrary) and $\lambda_+= 0.302$, $\omega= 0.074$ (determined by minimizing $c_6$). Darker colors are smaller values and lighter colors are higher values of the cost function. The blue canyon shaped region is where $c_{43}$ is close to its minimum.}
		\end{center}
	\end{figure}

	After determining $\lambda_+$ and $\omega$, we used the deaths data up to day 43 to try to determine the parameter $\xi$ and the reducing factor $\epsilon_1$ for the infection rate during the social distancing measures started at day 6. Similar to Lombardy, Fig. \ref{spquarant} shows a canyon shaped region in which the cost function $c_{43}$ is close to its minimum. The parameters used in the figure, besides the optimal values for $\lambda_+$ and $\omega$, are specified in its caption.
	
\begin{figure} 
		\begin{center}
			\includegraphics[width= 0.9\textwidth]{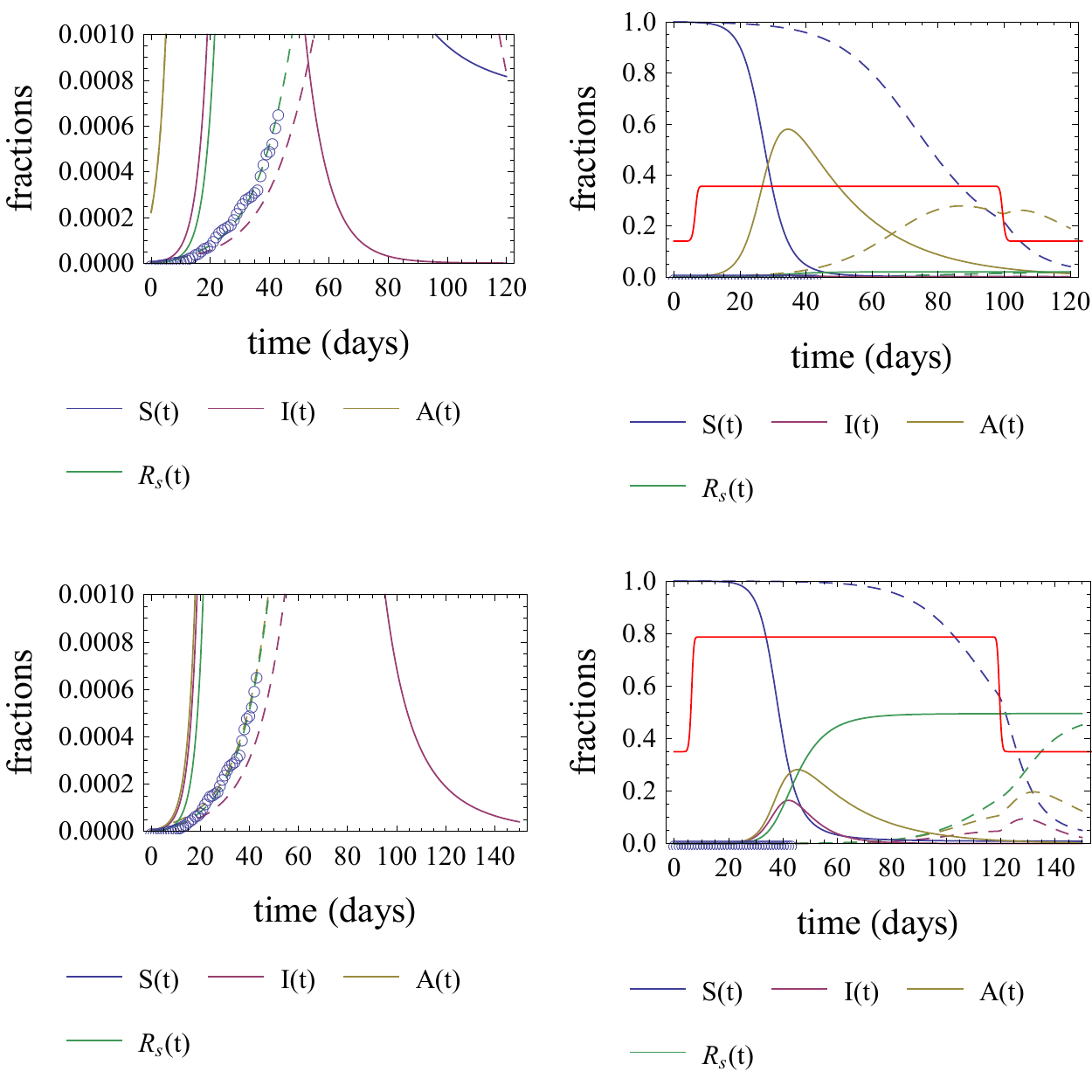}
			\caption{\label{finalsp} Plots of expected outcomes for the COVID-19 epidemic in the state of S\~ao Paulo for up to 150 days after Mar. 17.	The blue dots compare the results with epidemic deaths data. The optimistic and pessimistic outcomes are presented in the first and second row, respectively. The plots in the different columns are different only in the range of the vertical scale. The parameter values valid for both rows are: $\gamma_s=1/7$, $\gamma_a=1/21$, $\lambda_+= 0.302$, $\omega=0.074$, $\mu=0.5$. For the first row only, $\epsilon_1=0.395$, $\xi=0.02$
				and the social isolation measures were relaxed at day 100. For the second row only, $\epsilon_1=0.445$, $\xi=0.5$ and the social isolation measures were relaxed only at day 120. The red lines appearing in the second column are the graphs of $\frac{\gamma_a}{\beta(t) (1-\xi) \mu}$, see Eq.~(\ref{approxcond}). 
				Notice that with a good approximation, when the graph of $S(t)$ is below the red line, the fraction of MSA infected individuals decreases.}
		\end{center}
	\end{figure}

	As in the case of Lombardy, we are faced with the fact that good fits of the A-SIR model with the deaths data can be obtained for a continuum of values for $\xi$. For small values of $\xi$ we have optimistic outcomes, in the sense that the number of deaths at the end of the epidemic is smaller. For larger values of $\xi$, we have pessimistic scenarios. Fig. \ref{finalsp} shows results both of an optimistic and a pessimistic solution, with different vertical scales.
	
	One important difference between the cases of S\~ao Paulo and Lombardy is that the social distancing period started much earlier in S\~ao Paulo. As a consequence, the fraction of 
	deceased people in S\~ao Paulo is much smaller up to now. The data points in the graphs of the second column in Fig. \ref{finalsp} are almost invisible. Of course, this is good, but it also has a bad side. In both optimistic and pessimistic cases the fraction of susceptible individuals remains for a long period 
	above the threshold in the right-hand side of Eq.~(\ref{approxcond}). Consequently, the fraction of infected individuals will be growing for a longer time when compared to Lombardy. 
	
	In the optimistic case, we see that if the social distancing measures remain with the present intensity until day 100 (Jun. 23), the fraction of infected individuals will increase up to a maximum by day 80 and start decreasing. If the social distancing measures are removed at day 100, the fraction of infected individuals will resume growth for some days and then it will decrease again. 
	
	In the pessimistic case, the number of infected individuals will grow to a number much larger than the present one. Even if the social distancing measures are relaxed after day 120, there will still be a rapid increase in the number of infected people. 
	
	In both cases, we predict that the number of infected individuals at present 
	will still  increase considerably before attaining its peak value, which will be by day 90 in the optimistic case, or by day 130 in the pessimistic case. Given that the health services at S\~ao Paulo are already operating close to their maximum capacity, in both cases we see that it is necessary that social distancing is intensified in order to raise the threshold for the decrease of the infected fraction and prevent their collapse.
	
	\section{Conclusions}
	\label{secconc}
	The COVID-19 pandemic forced us, scientists, to tackle the difficult task of trying to understand a new disease at the same time it is killing people in our neighborhoods and stressing our health services. As shown in this paper, the lack of solid knowledge on basic questions produces also an ignorance of what 
	may happen in the future, even the present situation being well described by a simple mathematical model. We hope that more basic research may help fill the knowledge gaps, but probably that will take time.
	
	Since the beginning of the pandemic, we have seen many papers with very different predictions on the population fraction that may die as a consequence of COVID-19, most of them too catastrophic. Part of this disparity
	is due to our ignorance on basic facts about the virus and the disease, as already remarked, 
	and on the number of mildly symptomatic or asymptomatic cases. Another part is a consequence of the difficulty in estimating the many parameters in any realistic mathematical model.
	
	In this paper we used a model as simple as possible in order to have the minimum number of parameters, and devised a procedure to estimate these parameters based only on the more faithful data, the number of deaths. We avoided using information on the number of 
	confirmed cases, or trying to guess underreporting factors. If we had considered more complete models, we would probably have to estimate a larger number of parameters, resulting on a larger uncertainty. We believe that our results, uncertain as they are, may be useful in showing both a worst and a best possible outcome. One important result of our calculations is that the epidemics both in Lombardy and in S\~ao Paulo would have been much worse if social distancing measures had not been taken.
	
	We saw that in an optimistic possibility, the number of cases of COVID-19 in Lombardy 
	might not increase after the social distancing measures are relaxed. However,
	we also saw that it is likely
	that the number of cases shows a new quick rise,
	being then necessary either to keep these measures for longer, or use 
	alternative measures. 
	As social distancing is being relaxed at many countries, particularly in Italy, it is possible that the uncertainty in our results will be solved in the next days according to whether the number of cases will grow rapidly, or not.
	
	In S\~ao Paulo state, as the fraction of infected individuals is up to now much smaller than in Lombardy, herd immunity is still 
	far. As a consequence, our calculations 
	predict that the fraction of infected individuals will still increase for 
	some time, even in the most optimistic case. As stressed before, strengthening social distance measures could alleviate this situation.

	We believe that the two locations 
	in which we fitted the model to real data may serve as examples
	of what may happen in other locations.
	
	\section{Post scriptum} \label{secps}
	As this paper was being written, more data became available both for Lombardy and for S\~ao Paulo state. The paper was finished on May 21, but in all figures of Sect.~\ref{secfit} we decided to keep the data only up to Apr. 29, because the conclusions would not change too much. We report in this section some differences.
	
	The strict lockdown in Italy ended on May 4, but social isolation measures are being slowly relaxed. If we had added the most recent data to Fig.~\ref{finallomb}, we would still have good fits of the data to the model in both cases,
	suggesting that the infection rate has not yet increased very much. 
	
	If in Eq.~(\ref{DRs}) we use $t=85$ and the datum for the population fraction of deceased individuals up to that day, we obtain an estimate $R_s(85)=0.0306$. 
	Taking into account Eq.~(\ref{asympRs}) and the fact that $R_s$ is an increasing function, this means that although values of $\xi$ smaller than $0.0306$ are allowed by Fig.~\ref{lombquarant1}, these must be ruled out. The optimistic value $\xi=0.04$ used in Fig.~\ref{finallomb} remains as a possibility. 
	
	On May 13 appeared the first results of a serological study for a large random sample of the population in Spain \cite{spain}. Although we have not used any data of Spain in this paper, one result in the cited report is interesting to consider here. On page 12 of \cite{spain} a map shows 
	the percentages of people having antibodies against SARS-COV-2 in all provinces of Spain. In the provinces where the epidemic was stronger, these percentages vary between $10.9\%$ and $14.2\%$.  This means that the fraction of susceptible individuals varies between $0.858$ and $0.891$ in these provinces. Of course these susceptible fractions cannot be blindly extended to any other location, but if we extrapolate them to Lombardy, that would also rule out the the optimistic value, $\xi=0.04$, because it produces values for the susceptible fraction considerably smaller. 
	
	Regarding the epidemic in S\~ao Paulo,~Table \ref{tabsp} shows 
	the ratio of the maximum predicted symptomatic infected individuals to 
	the present value (day 65) of the same quantity
	in three situations: 
	the pessimistic and optimistic situations considered in Fig.~\ref{finalsp},
	and an intermediate situation (graphs not shown) with $\xi=0.12$, $\epsilon=0.405$ 
	and social distance measures with the present intensity up to day 120. 
	The table also shows the predicted date of the maximum.
	
	\begin{table}
		\begin{center}      
			\begin{tabular}{ccc}
				\hline\noalign{\smallskip}
				& Day/Date & $I_{max}/I(65)$  \\
				\hline\noalign{\smallskip} 
				Optimistic& 80 / June 5 & 1.38 \\
				Pessimistic& 130 / July 25 & 39.8\\
				Intermediate& 100/ June 25 & 5.49\\
				\noalign{\smallskip}\hline
			\end{tabular}
			\caption{Predicted day/date for the maximum of symptomatic infected individuals in the state of S\~ao Paulo and predicted ratio of the maximum symptomatic infected individuals to the present number of symptomatic infected individuals, day 65 (May 21). We consider three cases: the optimistic and pessimistic cases of Fig.~\ref{finalsp} and an intermediate one with $\xi=0.12$, $\epsilon=0.405$ and social distancing lasting up to day 120.\label{tabsp}}
		\end{center}
	\end{table}
	
	We see that even in an optimistic situation, an increase of $38\%$ in the number of cases in only 16 days is expected from the results of the A-SIR model. Intensified social distancing measures might help mitigating this situation. In the other two cases, the recommendation for intensifying social distancing is of course still stronger.
	
	\textbf{Acknowledgments:} We thank the members of the COVID-19 Modeling Task Force in Minas Gerais, Brazil, for suggesting many references and for discussions of results in earlier phases of this research.
	
	\textbf{Funding:} This research did not receive any specific grant from funding agencies in the public, commercial, or
	not-for-profit sectors.

\end{document}